\documentclass[prl,amsmath,twocolumn, showpacs, superscriptaddress,10pt]{revtex4-1}

\usepackage{amsmath}
\usepackage{hyperref}
\usepackage{graphicx}

\usepackage{color}
\definecolor{Blue}{rgb}{0.00, 0.00, 1.00}
\definecolor{Red}{rgb}{1.00, 0.00, 0.00}

\newcommand{\be}{\begin{equation}}
\newcommand{\ee}{\end{equation}}
\newcommand{\bea}{\begin{eqnarray}}
\newcommand{\eea}{\end{eqnarray}}

\begin{document}

\title{Finite temperature free fermions and the Kardar-Parisi-Zhang equation at finite time}

\author{David S. \surname{Dean}}
\affiliation{Univ. Bordeaux and CNRS, Laboratoire Ondes et Mati\`ere  d'Aquitaine
(LOMA), UMR 5798, F-33400 Talence, France}
\author{Pierre Le Doussal}
\affiliation{CNRS-Laboratoire de Physique Th\'eorique de l'Ecole Normale Sup\'erieure, 24 rue Lhomond, 75231 Paris Cedex, France}
\author{Satya N. \surname{Majumdar}}
\affiliation{Univ. Paris-Sud, CNRS, LPTMS, UMR 8626, Orsay F-91405, France}
\author{Gr\'egory \surname{Schehr}}
\affiliation{Univ. Paris-Sud, CNRS, LPTMS, UMR 8626, Orsay F-91405, France}

\begin{abstract} We consider the system of $N$ one-dimensional free fermions confined by a harmonic well  $V(x) = m\omega^2 {x^2}/{2}$ at finite inverse temperature $\beta = 1/T$. The average density of fermions  $\rho_N(x,T)$ at position $x$ is derived. For $N \gg 1$ and $\beta \sim {\cal O}(1/N)$, $\rho_N(x,T)$  is given by a scaling function interpolating between a Gaussian at high temperature, for $\beta \ll 1/N$, and the Wigner semi-circle law at low temperature, for $\beta \gg N^{-1}$. In the latter regime, we unveil a scaling limit, for $\beta {\hbar \omega}= b N^{-1/3}$, where the fluctuations close to the edge of the support, at $x \sim \pm \sqrt{2\hbar N/(m\omega)}$, are described by a limiting kernel $K^{\rm ff}_b(s,s')$ that depends continuously on $b$ and is a generalization of the Airy kernel, found in the Gaussian Unitary Ensemble of random matrices. Remarkably, exactly the same kernel $K^{\rm ff}_b(s,s')$ arises in the exact solution of the Kardar-Parisi-Zhang (KPZ) equation in 1+1 dimensions at finite time $t$, with the correspondence {$t= b^3$}. \end{abstract}

\maketitle


There is currently intense activity in the field of low dimensional quantum systems. This is motivated by new experimental developments for manipulating fundamental quantum systems, notably ultra-cold atoms \cite{BDZ08,GPS08}, where the confining potentials are optically generated. One of the most fundamental quantum systems is that of $N$ non-interacting spinless fermions in one dimension, confined in a harmonic trap $V(x) = \frac{1}{2} m\omega^2 x^2$. Recently, the zero temperature 
($T=0$) properties of this system have been extensively 
studied~\cite{GPS08,calabrese_prl,vicari_pra,vicari_pra2,vicari_pra3,eisler_prl,marino_prl,CDM14}, 
and a deep connection between this free fermion problem and random matrix theory (RMT) has been established. Specifically, the probability density function (PDF) of the positions $x_i$'s of the $N$ fermions, given by the modulus squared of the ground state wave function $\Psi_0(x_1, \cdots, x_N)$, can be  written as
\begin{equation}\label{psi0}
|\Psi_0(x_1, \cdots, x_N)|^2 = \frac{1}{z_N(\alpha)} \prod_{i<j} (x_i - x_j)^2 e^{-\alpha^2\sum_{i=1}^N x_i^2}
\end{equation}
with $\alpha = \sqrt{m\omega/\hbar}$, and  where $z_N(\alpha)$ is a normalization constant. Eq. (\ref{psi0}) shows that the rescaled positions $\alpha x_i$'s behave statistically as the  eigenvalues of random $N \times N$ matrices of the Gaussian
Unitary Ensemble (GUE) of random matrix theory (RMT) \cite{mehta,forrester}. 

In particular, the average density of free fermions $\rho_N(x,T=0)$ in the ground state 
is given, in the large $N$ limit, by the Wigner semi-circle law \cite{mehta,forrester}, 
\begin{equation}
\rho_N(x, T=0) \sim \frac{\alpha}{\sqrt{N}} \rho_{\rm sc}\left(\frac{\alpha \, x}{\sqrt{N}} \right) \;,
\label{scaling.1}
\end{equation}
where $\rho_{\rm sc}(x) = \frac{1}{\pi} \sqrt{2-x^2}$, on the finite support $[-\sqrt{2N}/\alpha, 
\sqrt{2N}/\alpha]$. An important property of free fermions at $T=0$, inherited from their connection with  the eigenvalues of random matrices (\ref{psi0}), is that they constitute a determinantal point process, for  any finite $N$. This means that their statistical properties are fully encoded in a two-point kernel $K_N(x,y)$ from which  any $k$-point correlation function can be written as a $k \times k$ determinant built from $K_N(x,y)$ [see Eqs. 
(\ref{def_correl}, \ref{determinantal}) below].

\begin{figure}[hh]
\includegraphics[width=0.9\linewidth]{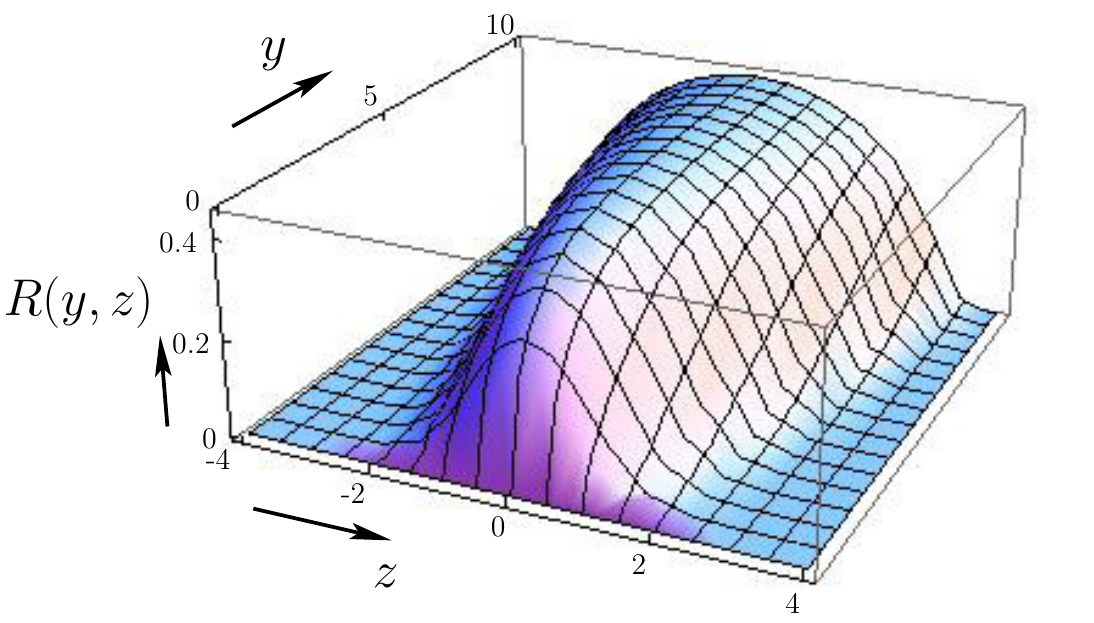}
\caption{(Color online) Plot of the scaling function $R(y,z)$ associated with the density $\rho_N(x,T)$ (\ref{dsc.1}), given in Eq.~(\ref{fyz}).}\label{fig_density_bulk}
\end{figure}

Early developments in RMT focused on the {\it bulk regime}, where both $x$ and $y$ are in the middle of the {\em Wigner sea}, say close to the origin where both $x \propto u/\sqrt{N}$ and $y\propto v/\sqrt{N}$ ({\em i.e}., of the order of the typical inter-particle spacing $\sim 1/(\rho_{\rm sc}(0) \sqrt{N})$). In this  region the statistics of eigenvalues, and consequently the positions of the free fermions at $T=0$  (\ref{psi0}), are described by the so called sine-kernel $K_N(x,y) \propto K_{\rm Sine}(u-v)$ where $K_{\rm 
Sine}(w) = \sin{(\pi w)}/(\pi w)$. More recently, there has been a huge interest in the statistics of  eigenvalues at the edge of the Wigner sea which, for fermions, corresponds to the fluctuations close to $x = \pm \sqrt{2N}/\alpha$. To probe these fluctuations at the edge, a natural observable is the position of  the rightmost fermion, $x_{\max}(T=0) = \max_{1\leq i \leq N} x_i$ in the ground state. From Eq. (\ref{psi0}) we can immediately infer that the typical (quantum) fluctuations of $x_{\max}(T=0)$, correctly shifted and scaled, are described by the Tracy-Widom (TW) distribution $F_2(\xi)$ associated with the 
fluctuations of the largest eigenvalue $\lambda_{\max}$ of GUE random matrices \cite{TW_GUE}. Namely, one  has
\begin{eqnarray}
\alpha \, x_{\max}(T=0) = \sqrt{2N} + \frac{1}{\sqrt{2}N^{1/6}} \chi_2 \;,
\end{eqnarray}
where $\chi_2$ is a random variable whose cumulative distribution $F_2(\xi)$ can be written as a Fredholm 
determinant~\cite{fredholm}
\begin{eqnarray}\label{eq:F2}
F_2(\xi) = \Pr[\chi_2 \leq \xi] = \det(I - P_\xi K_{\rm Ai} P_\xi) \;,
\end{eqnarray} 
where $K_{\rm Ai}(x,y)$ is the  Airy kernel \cite{TW_GUE,For93}
\begin{eqnarray}\label{eq:Airy}
K_{\rm Ai}(x,y) = \frac{{\rm Ai}(x) {\rm Ai}'(y) - {\rm Ai}'(y) {\rm Ai}(x)}{x-y} \;.
\end{eqnarray}
Here ${\rm Ai}(x)$ is the Airy function and $P_\xi$ is the projector on the interval $[\xi,+\infty)$. 
Remarkably, it was found \cite{eisler_prl} that for a generic confining potential of the form $V(x) \sim 
x^p/p$, {the local fluctuations in the fermion problem at $T=0$ are universal both in the bulk and at the edge (\ref{eq:Airy}).} 


{Given this beautiful connection between free fermions in a 
harmonic trap at $T=0$ and RMT (\ref{psi0}), it is natural    
to study the effect of 
non-zero temperature in the free fermion system, for which much less is known (see however 
\cite{MNS94, Verba})}. Here, we analyze
this system at finite $T$ and find a very rich behavior 
for the {{average}} density of fermions, as well as for the fluctuations of the {right-most fermion}. In particular, we find a fascinating link between free fermions at {\em finite} $T$ and the Kardar-Parisi-Zhang (KPZ) equation in $1+1$ dimensions at {\em finite time}.

{{\it Summary of results.} First we compute the average density $\rho_N(x,T)$, for large $N$. We find that there  are  two natural dimensionless scaling variables in this problem
\begin{eqnarray}
y =\beta\, N\, \hbar\, \omega \; , \; z =  x\, \sqrt{\frac{\beta}{2}\, m\, \omega^2} \label{sc.2} \;,
\end{eqnarray}  
in terms of which $\rho_N(x,T)$ takes the scaling form
\begin{equation}
\rho_N(x,T) \sim \frac{\alpha}{\sqrt{N}} R \left(\beta\, 
N\, 
\hbar\, \omega=y, x\,\sqrt{\frac{\beta}{2}\, m\omega^2}=z\right),
\label{dsc.1}
\end{equation}
which holds in the scaling limit: (a) $\beta\to 0$, $N\to \infty$
but keeping $y=\beta\, N\, \hbar\, \omega $ fixed (i.e., $T \sim N$) and (b)
$\beta\to 0$, $x\to \infty$ but keeping $z=x\, \sqrt{\frac{\beta}{2}\, m\, 
\omega^2}$ fixed (i.e., $x \sim \sqrt{T}$).  We find that the scaling function $R(y,z)$, for all $y$ and $z$, is given by
\begin{equation}
R(y,z)= -\frac{1}{\sqrt{2\pi\,y}}\, 
{\rm Li}_{1/2}\left(-\left(e^y-1\right)\, 
e^{-z^2}\right) \;,
\label{fyz}
\end{equation}
where ${\rm Li}_n(x)=\sum_{k=1}^{\infty} x^k/k^n$ is the polylogarithm function. In Fig. \ref{fig_density_bulk} we
show a 3d-plot of $R(y,z)$. We can check, from an asymptotic analysis of $R(y,z)$ in (\ref{fyz}), that Eq. (\ref{dsc.1}) interpolates between the Wigner semi-circle (\ref{scaling.1}) in the limit $T \to 0$
and the classical Gibbs-Botlzmann distribution for $T \to \infty$:
\begin{equation}
\rho_N(x,T\to \infty) \sim \sqrt{\frac{\beta\,m\, \omega^2}{2\pi}}\, 
\exp\left[-\frac{\beta}{2}\, m\,\omega^2\, x^2\right]\, \;,
\label{Tinf}
\end{equation}
which holds also in the scaling limit $\beta\to 0$, $x\to \infty$ but 
keeping $x\,\sqrt{\beta}$ fixed (with the limit $N\to \infty$ already 
taken). Note that the physical mechanism behind this interpolation is {{very}}
different from {{those}} found earlier in other matrix models~\cite{JP1,JP2}.}

We {{next}} consider a different {\em low temperature} scaling limit where $T = b^{-1} N^{1/3} \hbar \omega$, 
corresponding to the case  ${{y\to \infty}}$ in Eq. (\ref{dsc.1}). In this scaling limit $\rho_N(x,T = {b}^{-1} N^{1/3}\hbar \omega)$ 
is thus given by the Wigner semicircle (\ref{scaling.1}), which has a finite support $[-\sqrt{2N}/\alpha, \sqrt{2 N}/\alpha]$. We show that for $N \gg 1$, $N$ free fermions at finite temperature in the canonical ensemble behave asymptotically as a determinantal point-process (which is not true for finite $N$). 
{A similar process was studied in \cite{Joh07}, in the context of the matrix model introduced in Ref. \cite{MNS94}.}

Close to the edge where $x \approx \sqrt{2N}/\alpha$, and for $\beta \hbar \omega = b N^{-1/3}$ we show that this determinantal point process is characterized by a limiting kernel $K^{\rm ff}_{b}(s,s') $ given by
\begin{eqnarray}\label{kff}
K^{\rm ff}_{b}(s,s') = \int_{-\infty}^\infty \frac{[{\rm Ai}(s+u){\rm Ai}(s'+u) ]}{e^{-b\, u} +1} du \;,
\end{eqnarray}
which is a generalization of the Airy-kernel (\ref{eq:Airy}). As a consequence, we find that the cumulative distribution of the position
of the rightmost fermion $x_{\max}(T)$ is given by the following Fredholm determinant \cite{fredholm}
\begin{equation}\label{gap_proba}
\Pr\left(x_{\max}(T) \leq \frac{\sqrt{2N}}{\alpha} + \frac{N^{-\frac{1}{6}}}{\alpha \sqrt{2}}\xi\right) 
\underset{N \to \infty}{\to} \det(I - P_\xi K^{\rm ff}_{b} P_\xi).
\end{equation}
Note that, using that $\lim_{b \to \infty} K^{\rm ff}_{b}(s,s') = K_{\rm Ai}(s,s')$ in Eq.~(\ref{gap_proba}), we recover the TW distribution of Eq. (\ref{eq:F2}) in the limit $b \to \infty$, as one should. 

Remarkably, exactly the same expression as Eq. (\ref{gap_proba}) was recently found in the 
study of the $(1+1)$-d KPZ equation
in curved geometry. The KPZ equation describes the time evolution 
of a height field $h(x,t)$ at point $x$ and time $t$ {as follows
\begin{eqnarray}\label{eq:KPZ}
\partial_t h = \nu \partial_x^2 h + \frac{\lambda_0}{2}\, (\partial_x h)^2 + \sqrt{D} \eta(x,t) \;,
\end{eqnarray}
where} $\eta(x,t)$ is a Gaussian white noise with zero mean and 
{correlator} 
$\langle \eta(x,t) \eta(x',t')\rangle = \delta(x-x')\delta(t-t')$.
We start from the narrow wedge initial condition, $h(x,0) = - |x|/\delta$, with $\delta \ll 1$,
which gives rise to a curved (or {\em droplet}) {mean profile} as time evolves~\cite{ISS13}.
{Defining the natural time unit $t^*=2(2 \nu)^5/(D^2 \lambda_0^4)$ and $\gamma_t=(t/t^*)^{1/3}$ \cite{footnote}, the
time-dependent generating
function 
\begin{eqnarray}
g_{t}(\zeta) = \langle \exp(- e^{\gamma_t (\tilde h(0,t) - \zeta)})\rangle \;
\end{eqnarray}
of $\tilde h(0,t) = (\frac{\lambda_0 h(0,t)}{2 \nu} + \frac{t}{12 t^*})/\gamma_t$, the 
rescaled height at $x=0$,  
is expressed as a Fredholm determinant \cite{ISS13,SS10,CLR10,DOT10,ACQ11}:
\begin{eqnarray}
&&g_{t}(\zeta) = \det(I - P_\zeta K^{\rm KPZ}_{t} P_\zeta) \label{GF}\\
&&K^{\rm KPZ}_{t}(x,y) = \int_{-\infty}^\infty \frac{{\rm Ai}(z+x) {\rm Ai}(z+y)}{e^{-\gamma_t z} + 1} 
\,{dz}  \label{eq:Airy_def} \;.
\end{eqnarray}
Comparing Eqs. (\ref{kff}) for the fermions and (\ref{eq:Airy_def}) for KPZ, we see that the two kernels 
are the same $K^{\rm ff}_{b} = K^{\rm KPZ}_{{{t = t^*\,b^3}}}$ in time unit $t^*$.} While comparison of Eqs. 
(\ref{gap_proba}) and (\ref{GF}) show that the cumulative distribution of $x_{\max}(T)$ for the free 
fermion problem is the same as the generating function {$g_{{{t=t^*\,b^3}}}(\zeta)$ in the KPZ equation.}
{One can show that it interpolates between a Gumbel distribution at high $T$ (small time $t$ for KPZ) where the fermions are uncorrelated, to the Tracy-Widom distribution at $T \to 0$ (large time~$t$ for KPZ).}


{\it Average density of fermions.} The joint probability density function (PDF) of the 
positions of the $N$ fermions is constructed from the single particle wave functions 
\begin{eqnarray}
\varphi_k(x) = \left[\frac{\alpha}{\sqrt{\pi} 2^k k!}\right]^{1/2} e^{-\frac{\alpha^2 x^2}{2}} H_k(\alpha x) \;,
\end{eqnarray}
where $H_k$ is the Hermite polynomial of degree $k$. In the canonical ensemble it  
is given by the  Boltzmann weighted sum of slater determinants (see \cite{unp} for details)
\begin{eqnarray}\label{p_start}
P_{\rm joint}(x_1, \cdots x_N) &=& \frac{1}{{N!}Z_N(\beta)}\sum_{k_1< \cdots < k_N} \left[\det_{1\leq i,j \leq N} (\varphi_{k_i}(x_j)) \right]^2 \nonumber \\
&\times& e^{-\beta(\epsilon_{k_1}+\cdots+ \epsilon_{k_N})} \;,
\end{eqnarray}
where $Z_N(\beta)$ is a normalization constant and where $\epsilon_k = \hbar\omega(k+1/2)$'s are single particle energy 
levels (in Eq. (\ref{p_start}) all the $k_i$'s range from $0$ to $\infty$).  We first  compute the mean density of free fermions $\rho_{N}(x,T) = N^{-1} \sum_{i=1}^N \langle \delta(x-x_i)\rangle$, where $\langle \ldots \rangle$ means an average computed with (\ref{p_start}). This amounts, up to a multiplicative constant, to integrating 
the joint PDF $P_{\rm joint}(x,x_2, \cdots x_N)$ over the last $N-1$ variables, yielding a rather complicated expression 
which, however, simplifies in the large $N$ limit where the canonical ensemble and the grand-canonical ensemble become 
equivalent~\cite{unp}. Hence, for large $N$ one obtains (see also \cite{MNS94, Verba})
\begin{eqnarray}
\rho_{N}(x,T) \approx \frac{1}{N} \sum_{k=0}^\infty \frac{[\varphi_k(x)]^2}{e^{\beta(\epsilon_k - \mu)} + 1} \label{eq_rho}\;, 
\end{eqnarray}
where $1/(e^{\beta(\epsilon_k-\mu)}+1)$ is the Fermi factor and the chemical potential $\mu$ is fixed by imposing that mean number of fermions is  $N$:
\begin{eqnarray}
N = \sum_{k=0}^\infty \frac{1}{e^{\beta(\epsilon_k - \mu)} + 1} \label{eq_N} \;.
\end{eqnarray}  
We first analyze Eqs. (\ref{eq_rho}) and (\ref{eq_N}) in the scaling limit where, $N \to \infty$, with  $\beta \to 0$ and 
$x \to \infty$ but keeping   $y$ and $z$ defined in (\ref{sc.2}) fixed (implying in particular $\beta \sim 1/N$). In this limit, the sums in (\ref{eq_rho}) and (\ref{eq_N}) can be replaced by integrals and, from (\ref{eq_N}), we find {$e^{\beta \mu} = e^{y}-1$}. Then using 
the asymptotic behavior of the Hermite polynomials $H_n(x)$ for large degree $n$, one obtains~\cite{unp} 
the scaling form {{in}} Eq. (\ref{dsc.1}) together with the explicit expression of the scaling function in (\ref{fyz}). 

\begin{figure}
\includegraphics[width=0.95\linewidth]{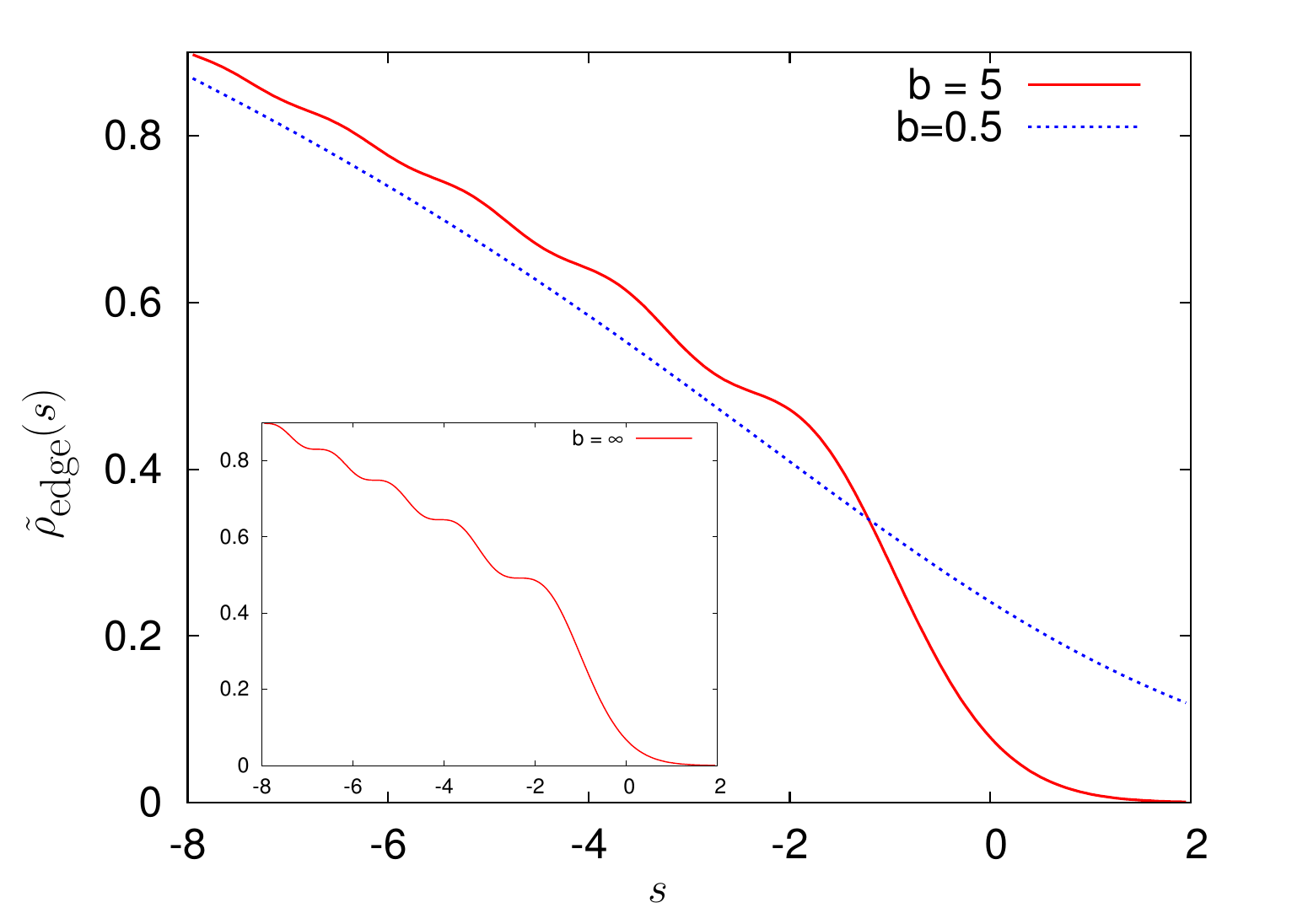}
\caption{(Color online) Plot of $\tilde \rho_{\rm edge}(s)$ given in Eq. (\ref{expr_edge_rho}) corresponding to two different (scaled) temperatures $b=0.5$ (dotted line) and $b=5$ (solid line). {\bf Inset:} plot of $\tilde \rho_{\rm edge}(s)$ corresponding to $b \to \infty$ given in the text and shown here for comparison with the main plot.}\label{fig_density_edge}
\end{figure}

We {{then}} turn to Eqs. (\ref{eq_rho}) and (\ref{eq_N}) 
in the scaling limit where $\beta {\hbar \omega} = b N^{-1/3}$, thus $y \to \infty$. Hence the average density is 
given by its, Wigner-semi-circle, $T=0$ limit (\ref{scaling.1}). In this regime, an interesting scaling limit emerges 
close to the edges, for $x =\pm \sqrt{2N}/\alpha$.   
To analyze the behavior of $\rho_{N}(x,T)$ close
to $x = \sqrt{2N}/\alpha$ \cite{compm} we insert the expression of $\beta \mu \sim bN^{2/3}$, 
obtained from (\ref{eq_N}), into (\ref{eq_rho})
and 
%
perform a change of variable in the sum, by setting $k = N + m$, to obtain:
\begin{eqnarray}\label{rho_N_inter}
\rho_N(x,T) \sim \frac{1}{N} \sum_{m=-N}^\infty \frac{[\varphi_{N+m}(x)]^2}{\exp{(b m/N^{1/3})}+1} \;.
\end{eqnarray} 
Using the Plancherel-Rotach formula for Hermite polynomials at the edge (see for instance Ref. \cite{FFG06}) yields:
\begin{equation}\label{Plancherel_Airy}
\varphi_{N+m}\left(\frac{\sqrt{2N}}{\alpha} + \frac{s}{\sqrt{2}\alpha} N^{-\frac{1}{6}} \right) \sim {\sqrt{\alpha}} \frac{2^{\frac{1}{4}}}{N^{\frac{1}{12}}} {\rm Ai}\left(s - \frac{m}{N^{\frac{1}{3}}} \right)  \;,
\end{equation} 
up to terms of order ${\cal O}(N^{-2/3})$. Hence by 
inserting this asymptotic formula (\ref{Plancherel_Airy}) into Eq. (\ref{rho_N_inter}) and replacing the discrete sum over $m \sim N^{1/3}$ by an integral we obtain:
%
\begin{equation}
\rho_{N}\left(\frac{\sqrt{2N}}{\alpha} + \frac{s}{\sqrt{2}\alpha} N^{-\frac{1}{6}},b^{-1}N^{\frac{1}{3}} \hbar \omega\right)  \sim {\alpha} N^{-\frac{5}{6}} \tilde \rho_{\rm edge}(s) \;,
\end{equation}
where $\tilde \rho_{\rm edge}(s)$ is given by
\begin{eqnarray}\label{expr_edge_rho}
&&\tilde \rho_{\rm edge}(s) = \sqrt{2} K^{\rm ff}_{b}(s,s) \;,
\end{eqnarray}
and the kernel $K^{\rm ff}_{b}(s,s')$ is given in (\ref{kff}). In the zero-temperature limit $b \to \infty$, we recover  
$\tilde \rho_{\rm edge}(s) \sim \sqrt{2} \left[({\rm Ai}'(s))^2 -s {\rm Ai}^2(s)\right]$,
%
the standard result for the mean density of eigenvalues at the edge of the spectrum of GUE random matrices \cite{BB91, For93}. In Fig. \ref{fig_density_edge}, we show how $\tilde \rho_{\rm edge}(s)$ behaves for different values of the reduced inverse temperature $b$. 

\begin{figure}[h]
\includegraphics[width = 0.95\linewidth]{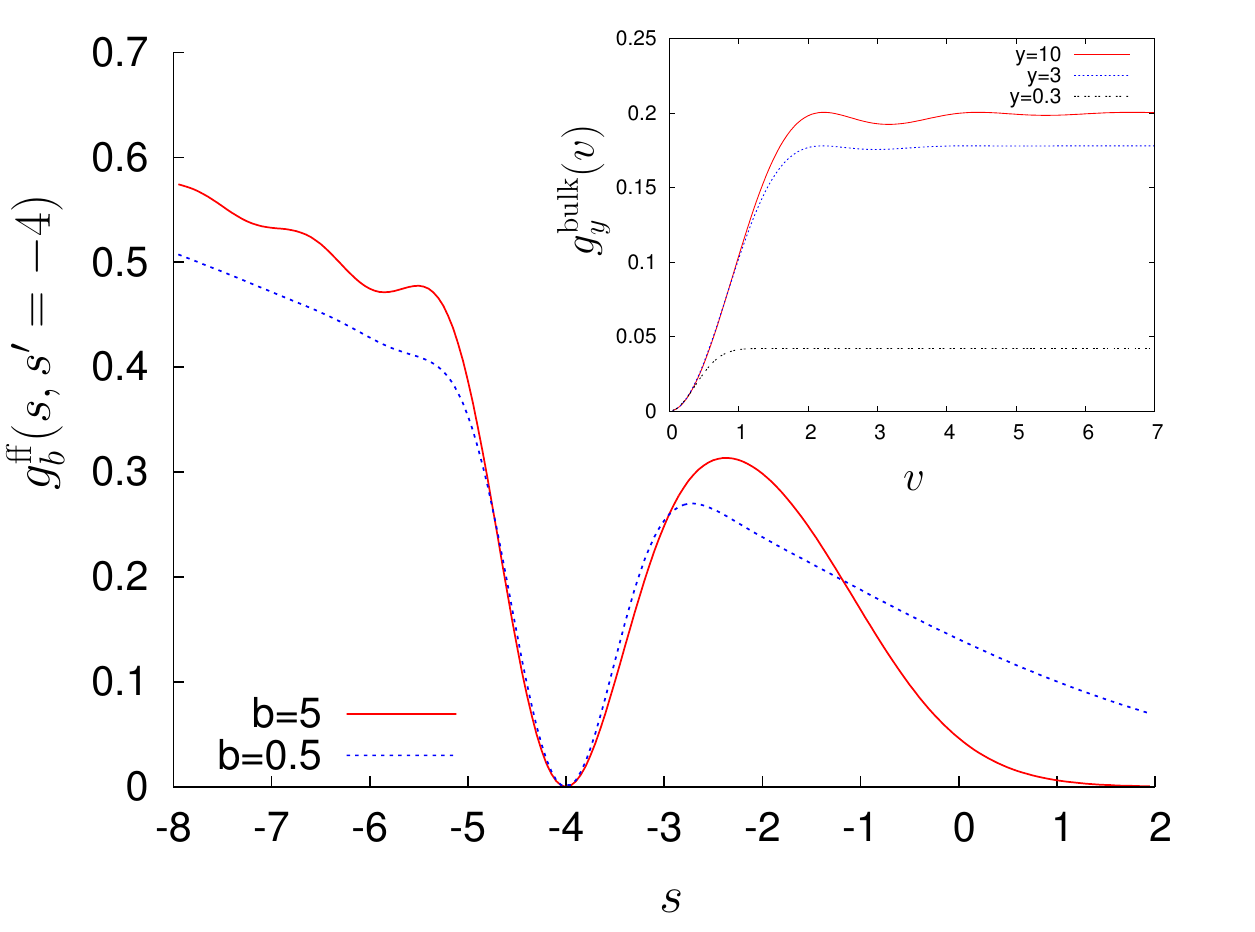}
\caption{(Color online) Plot of the 2-point correlation function at the edge $g^{\rm ff}_b(s,s'=-4)$ as a function of $s$ and for  scaled inverse temperatures $b=0.5$ and $5$. {\bf Inset:} Plot of the 2-point correlation function in the bulk $g^{\rm bulk}_y(v)$ versus $v$ for scaled inverse temperatures $y = 0.3, 3$ and $10$.} \label{fig_kernels}
\end{figure}

{\it Kernel and higher order correlation functions.} More generally, one can study the $n$-point correlation function for $N$ free fermions at finite temperature. We define $R_n(x_1,\cdots x_n)$ as
\begin{eqnarray}\label{def_correl}
&&R_n(x_1, \cdots, x_n) = \frac{N!}{(N-n)!} \int_{-\infty}^\infty dx_{n+1} \cdots \int_{-\infty}^\infty dx_{N} \nonumber \\
&&\times P_{\rm joint}(x_1, \cdots, x_n, x_{n+1}, \cdots, x_N)
\end{eqnarray}
where $P_{\rm joint}(x_1, \cdots, x_N)$ is the joint PDF of the $N$ fermions at finite temperature of Eq. (\ref{p_start}) \cite{comR}. Using the equivalence, in the large $N$ limit, between the canonical and grand-canonical ensembles, one can show that 
\begin{eqnarray}\label{determinantal}
R_n(x_1, \cdots, x_n) \approx \det_{1\leq i,j \leq n} K_N(x_i,x_j)
\end{eqnarray}
where the kernel $K_N(x,x')$ is given by
\begin{eqnarray}\label{expr_kernel}
K_N(x,x') = \sum_{k=0}^\infty \frac{\varphi_k(x)\varphi_k(x')}{e^{\beta(\epsilon_k - \mu)} + 1} \;.
\end{eqnarray}
We first analyze the kernel (\ref{expr_kernel}) in the scaling limit where $N \to \infty$ and $\beta \to 0$ keeping $y$ in (\ref{sc.2}) fixed (i.e., $\beta \sim {\cal O}(1/N)$). In this limit, if we are interested in the behavior of $K_N(x,x')$ in the 
{\em bulk} where both $x \sim (u/\alpha) N^{-1/2}$ and $x' \sim (u'/\alpha) N^{-1/2}$ are close to the origin, one finds (see also \cite{Verba, Joh07})
\begin{eqnarray}\label{K_bulk}
&&\lim_{N \to \infty}\frac{N^{-\frac{1}{2}}}{\alpha}K_N\left(\frac{u}{\alpha}N^{-\frac{1}{2}}, \frac{u'}{\alpha}N^{-1/2}\right) \sim K^{\rm bulk}_{y}\left(u-u'\right) \nonumber \\
&& K^{\rm bulk}_{y}(v) = \frac{1}{\pi \sqrt{2y}} \int_0^\infty \frac{\cos{\left(\sqrt{\frac{2 p}{y}} v\right)}}{(1+e^p/(e^y-1))\sqrt{p}} \, dp \;.
\end{eqnarray}  
Note that in the low temperature limit $y \to \infty$, the Fermi factor in Eq. (\ref{K_bulk}) behaves like a theta function, $\propto\theta(y-{p})$, implying that
$\lim_{y \to \infty} K^{\rm bulk}_{y}(v)/K^{\rm bulk}_{y}(0) = [\sin{(v \sqrt{2})}]/(v \sqrt{2})$, which (up to a scaling factor) is the expected sine-kernel. In the inset of Fig. \ref{fig_kernels} we show a plot of the pair-correlation function $g_y^{\rm bulk}(s) = [K^{\rm bulk}_y(0)]^2 - [K^{\rm bulk}_y(s)]^2$ for different values of the scaled inverse temperature $y$. 

However, in the low temperature scaling limit where $\beta {\hbar \omega}= b N^{-1/3}$, the kernel in the bulk is given by the sine-kernel while the interesting behavior occurs at the edge $x \sim \pm \sqrt{2N}/\alpha$. In this limit, performing the same analysis as above (\ref{rho_N_inter})-(\ref{expr_edge_rho}), one finds that the kernel $K_N(x,x')$ (\ref{expr_kernel}) takes the scaling form:
\begin{eqnarray}\label{largeN_edge}
\lim_{N \to \infty} \frac{N^{-\frac{1}{6}}}{\sqrt{2}\alpha} &&K_N\left(\frac{\sqrt{2N}}{\alpha} + \frac{s}{\sqrt{2}\alpha}N^{-\frac{1}{6}}, \frac{\sqrt{2N}}{\alpha} + \frac{s'}{\sqrt{2}\alpha}N^{-\frac{1}{6}}\right) \nonumber \\
= &&K_b^{\rm ff}(s,s') \;,
\end{eqnarray}
where $K_b^{\rm ff}(s,s')$ is given in Eq. (\ref{kff}). In Fig. \ref{fig_kernels} we show a plot of the 2-point correlation function at the edge $g^{\rm ff}_b(s,s') = K_b^{\rm ff}(s,s) K_b^{\rm ff}(s',s') - [K_b^{\rm ff}(s,s')]^2$ for different scaled inverse temperatures $b$. The properties of determinantal point processes {\cite{johansson}} then imply that the cumulative distribution of the position of the rightmost fermion $x_{\max}(T)$ is given by Eq.~(\ref{gap_proba}). 

We have analyzed the effect of finite temperature on 
free spinless fermions in a harmonic trap in one dimension. The scaling function, showing how the  average density of the system, in the bulk, crosses over from the Gaussian Gibbs-Boltzmann form at high temperatures (\ref{Tinf}) to the Wigner semi-circle law (\ref{scaling.1}) at low temperatures, has been computed. For large $N$, the equivalence between  canonical and grand canonical ensembles {implies} that free fermion statistics can be described as a determinantal process even at finite temperature. We derived the kernel for this process at finite temperature in the bulk, and also close to the edge of the bulk at low {temperature}. The statistics of the rightmost fermion {turns} out to be governed by a finite temperature generalization of the TW distribution. The temperature dependent kernel found here also {exhibit} a tantalizing connection with {the one} appearing in exact solutions of the KPZ equation \cite{SS10, CLR10, DOT10, ACQ11} at finite times $t$ with the correspondence $t=2b^3$. This connection in fact holds for  generic confining potentials such that $V(x) \sim x^p/p$, with $p>0$, for $x$ large: the kernel is universal, and only the scalings with $N$ are modified \cite{unpub}. 
{This intriguing connection between free fermions at finite temperature and KPZ at finite time is, for the moment, a pure observation, which still lacks a deeper physical understanding. Whether it is merely an accident (holding for droplet KPZ initial conditions and in $1d$), or the sign of a more fundamental connection 
will 
be interesting to explore in the future.}

\acknowledgments{We acknowledge stimulating discussions with P. Calabrese, T. Imamura, S. M. Nishigaki
and T. Sasamoto. We acknowledge support from PSL grant ANR-10-IDEX-0001-02-PSL
(PLD) ANR grant 2011-BS04-013-01 WALKMAT and in part by the Indo-French
Centre for the Promotion of Advanced Research under Project 4604-3 (SM and GS) and Labex-
PALM (Project Randmat) (GS).}

\newpage

\onecolumngrid

\begin{center}
{\Large Supplementary Material  \\ 
}
\end{center}

We give the principal details of the calculations described in the manuscript of the Letter.

\section{Average density}

The starting point of our computations is the joint probability density function (PDF) of the positions of the $N$ fermions at finite temperature $T = 1/\beta$. It can be expressed in terms of the single particle wave functions 
\begin{eqnarray}\label{def_phik}
\varphi_k(x) = \left[\frac{\alpha}{\sqrt{\pi} 2^k k!}\right]^{1/2} e^{-\frac{\alpha^2 x^2}{2}} H_k(\alpha x) \;,
\end{eqnarray}
where $H_k$ is the Hermite polynomial of degree $k$ and $\alpha=\sqrt{m \omega/\hbar}$. 
This single particle wavefunction has an associated energy eigenvalue, $\epsilon_k= \hbar\omega(k+1/2)$
where $k$ is an integer which ranges from $0$ to $\infty$. From these single particle states, one can
construct any generic many-body free fermion wavefunction by putting $N$ fermions in $N$ different
single particle levels indexed by $k_1<k_2<\ldots<k_N$. The fermionic nature 
of the particles allows at most one particle in a given single particle level. The energy of such a many-body state
is evidently, $E= \sum_{i=1}^N \epsilon_{k_i}=\hbar\omega\,(k_1+k_2+\dots+k_N+ N/2)$ and the many-body wavefunction 
is just the Slater determinant $\displaystyle{\Psi_E(\{x_i\})= 
\det_{1\leq i,j \leq N} (\varphi_{k_i}(x_j))}$. 
Assuming that the system is in the canonical ensemble characterized by the inverse temperature $\beta$, 
the Boltzmann weight associated with such an excited state is simply $e^{-\beta E}= e^{-\beta\,
\sum_{i=1}^N \epsilon_{k_i}}$.
Hence the
joint PDF of the particle positions in the canonical ensemble  
is given by the Boltzmann weighted sum of slater determinants
\begin{eqnarray}\label{p_start}
P_{\rm joint}(x_1, \cdots x_N) &=& \frac{1}{{N!}Z_N(\beta)}\sum_{k_1< \cdots < k_N} 
\left[\det_{1\leq i,j \leq N} (\varphi_{k_i}(x_j)) \right]^2 e^{-\beta\, (\epsilon_{k_1}+\cdots+ \epsilon_{k_N})} \;,
\end{eqnarray}
where $Z_N(\beta)$ is the canonical partition function of the free fermion gas
\begin{eqnarray}\label{partition_function}
Z_N(\beta) = \sum_{k_1<k_2<\ldots< k_N} e^{-\beta\, (\epsilon_{k_1}+\cdots+ \epsilon_{k_N}) } \;.
\end{eqnarray}
It is easy to check that $Z_N(\beta)$ is such that the PDF $P_{\rm joint}(x_1, \cdots x_N)$ is normalized to 
unity. 

We first compute the mean density of free fermions $\rho_{N}(x,T) = N^{-1} \sum_{i=1}^N \langle \delta(x-x_i)\rangle$, where $\langle \ldots \rangle$ means an average computed with (\ref{p_start}). This amounts, up to a multiplicative constant, to integrating 
the joint PDF $P_{\rm joint}(x,x_2, \cdots x_N)$ over the last $N-1$ variables. This yields
\begin{eqnarray}\label{start_rho}
\rho_N(x) = \frac{1}{N ! Z_N(\beta)} \sum_{k_1<k_2<\ldots< k_N} \int_{-\infty}^\infty dx_1  
\delta(x-x_1) \int_{-\infty}^\infty dx_2 \ldots   
\int_{-\infty}^\infty dx_N \left[\det_{1\leq i,j \leq N} 
(\varphi_{k_i}(x_j)) \right]^2 e^{-\beta(\epsilon_{k_1}+\cdots+ \epsilon_{k_N})} \;.
\end{eqnarray}

To make analytic progress with these formula in Eqs. (\ref{start_rho}) and (\ref{partition_function}), 
it is convenient to introduce the occupation numbers $m_k = 0,1$ which denote the number of fermions 
occupying the level $k$, of energy $\epsilon_k = \hbar \omega(k+1/2)$. In terms of these occupation numbers $m_k$'s, the partition function is given by
\begin{eqnarray}\label{partition_occupation}
Z_N(\beta) = \sum_{\{m_k\}} \left[ e^{-\beta\, \sum_{k\geq 0} m_k \epsilon_k} \delta\left(\sum_{k\geq0} m_k, N\right) \right] \;,
\end{eqnarray}
where $\sum_{\{m_k\}}$ denotes the sum over all the possible occupation numbers 
$m_k = 0,1$ for $k=0,1,2, \cdots$. In Eq. (\ref{partition_occupation}) 
$\delta(i,j) = 1$ if $i=j$ and $\delta(i,j) = 0$ if $i\neq j$: this Kronecker delta function thus 
imposes the total number of particles to be exactly $N$, as we are working in the canonical ensemble.
Consider the density $\rho_N(x)$ in Eq. (\ref{start_rho}). 
By writing out the Slater determinant explicitly, squaring it and integrating over $(N-1)$ coordinates
in Eq. (\ref{start_rho}) upon using the orthonormal properties of the single particle wavefunctions, 
it is not difficult to show that the density $\rho_N(x)$ in Eq. (\ref{start_rho}) can be written in terms of the $m_k$'s as
\begin{eqnarray}\label{rho_occupation}
\rho_N(x) =  \frac{1}{{Z_N(\beta)}} \frac{1}{N} \sum_{\{m_k\}} \left[  \left(\sum_{k\geq 0} m_k |\varphi_k(x)|^2 \right)  e^{-\beta \sum_{k\geq 0} m_k \epsilon_k} \delta\left(\sum_{k\geq 0} m_k,N\right) \right] \;.
\end{eqnarray}
Note that in the limit where $T \to 0$, the system is in the ground state characterized by 
$m_k = 1$ if $k=0,1,2, \cdots, N-1$ and $m_k = 0$ if $k \geq N$. Hence in this limit, Eq. (\ref{rho_occupation}) reads
\begin{eqnarray}
\rho_N(x) = \frac{1}{N} \sum_{k=0}^{N-1} |\varphi_k(x)|^2 \;,
\end{eqnarray}
as expected from the connection between free fermions at $T=0$ and random matrices belonging to the Gaussian Unitary Ensemble (GUE). What happens at finite temperature $T$ in the large $N$ limit? One can  show (see below) that in the limit of large $N$, one can replace the occupation numbers $m_k$ in Eqs. (\ref{partition_occupation}) and (\ref{rho_occupation}) by their average value $\langle m_k \rangle$ such that
\begin{eqnarray}\label{rho_N_grand_canonical}
\rho_N(x) \simeq \frac{1}{N} \sum_{k = 0}^\infty \langle m_k \rangle |\varphi_k(x)|^2 \;.\label{dens1}
\end{eqnarray}
In the thermodynamic language, this amounts to use the equivalence, in the large $N$ limit, between the canonical ensemble and the grand-canonical ensemble. The average value of the occupation number $\langle m_k \rangle$ is then given by the Fermi factor
\begin{eqnarray}\label{fermi_factor}
\langle m_k \rangle = \frac{1}{e^{\beta(\epsilon_k-\mu)} + 1} \;, \; \epsilon_k = \hbar \omega\left(k + \frac{1}{2} \right) \;, \; {\rm for} \; k = 0,1,2,\cdots \;.
\end{eqnarray}
In Eq. (\ref{fermi_factor}) the chemical potential $\mu$ is fixed by the total number of particles $N$, according to the relation
\begin{eqnarray}\label{chemical}
N = \sum_{k = 0}^\infty \langle m_k \rangle = \sum_{k = 0}^\infty \frac{1}{e^{\beta(\epsilon_k-\mu)}+1} \;.
\end{eqnarray}
 
We now analyze these formula (\ref{rho_N_grand_canonical})-(\ref{chemical}) in the bulk (the analysis at the edge is outlined in the Letter, see Eqs. (20)-(23)).

\subsection*{Asymptotic analysis in the bulk}

The bulk regime corresponds to the limit $\beta \to 0$, $N \to \infty$ and $x \to \infty$ keeping fixed the following dimensionless variables
\bea\label{def_yz}
y = \beta N \hbar \omega \quad , \quad z= x \sqrt{\beta m \omega^2/2} \;.
\eea 
In this limit, the equation fixing the chemical potential $\mu$ (\ref{chemical}) reads
\bea
&& N = \sum_{k=0}^\infty \frac{1}{e^{\beta (\hbar \omega (k+ \frac{1}{2})} e^{- \beta \mu} +1} 
\simeq \int dk \frac{1}{e^{k y/N} e^{- \beta \mu} +1} = \frac{N}{y} \ln(1+ e^{\beta \mu}) \;.
\eea 
This yields the relation
\bea\label{mu_bulk}
e^{\beta \mu} = e^y - 1 \;.
\eea
We now analyze the density $\rho_N(x)$ given in Eqs. (\ref{rho_N_grand_canonical}) and (\ref{fermi_factor}) which we evaluate for $x = z/\sqrt{\beta m \omega^2/2} = \sqrt{2N}/(\alpha \sqrt{y}) \gg 1$. After performing the change of variable 
$k = N p$ in the sum over $k$ in (\ref{rho_N_grand_canonical}), one obtains:
\bea\label{rho_bulk_inter}
&& \rho_N(x) \simeq \frac{1}{N} \sum_{k=0}^\infty \frac{\varphi_k(x)^2}{e^{\beta (\hbar \omega (k+ \frac{1}{2})} e^{- \beta \mu} +1}  
\simeq \int_0^\infty dp \frac{\left[\varphi_{N p}\left(x={z}\sqrt{2N}/(\alpha\, \sqrt{y}) \right)\right]^2}{e^{y p} (e^y -1)^{-1} +1}
\eea 
where we have replaced $\mu$ by its value given in Eq. (\ref{mu_bulk}) and where $\varphi_k(x)$ is given in Eq. (\ref{def_phik}). We now need an asymptotic expansion of $\varphi_k(x)$ for large $k$ (and large $x$). This expansion is provided by the Plancherel-Rotach formula [as given for instance in Eqs. (3.10) (3.11) of Ref. \cite{FFG06_supp}]. For $-1<X<1$, one has
\bea
&& e^{- M X^2} H_M(\sqrt{2 M} X) = \left(\frac{2}{\pi}\right)^{1/4} \frac{2^{M/2}}{(1-X^2)^{1/4}} M^{-1/4} (M!)^{1/2}  g_M(X) \left( 1 + {\cal O}\left(\frac{1}{N}\right) \right) \label{Plancherel_1}\\
&& {\rm with} \;\; g_M(X) = \cos\left( M X \sqrt{1-X^2} + (M + 1/2) \sin^{-1} X - M \pi/2\right)  \;. \label{Plancherel_2}
\eea 
Using these formulas (\ref{Plancherel_1}) and (\ref{Plancherel_2}) for $M=N p$ and $X=z/\sqrt{p y}$ -- see Eq. (\ref{rho_bulk_inter}) -- (taking into account that $X<1$, i.e., $p>z^2/y$) one finds
\bea
\rho_N\left(x = z/\sqrt{\beta m \omega^2/2} \right) &=& \frac{\alpha \sqrt{2}}{\pi} \int_{z^2/y}^\infty dp \frac{1}{e^{y p} (e^y -1)^{-1} +1} 
(N p)^{-1/2} \frac{1}{\sqrt{1- \frac{z^2}{p y}}} \left[g_{N p}(z/\sqrt{p y})\right]^2  \\
&=& \frac{\alpha \sqrt{2}}{\pi \sqrt{y} \sqrt{N}} \int_{z^2}^\infty dq \frac{1}{e^{q} (e^y -1)^{-1} +1} 
 \frac{1}{\sqrt{q - z^2}} \left[g_{N p}(z/\sqrt{q})\right]^2 \;, \label{rho_bulk_inter2}
\eea
where, in the second line, we have simply performed the change of variable $p = q/y$. To obtain the large $N$ limit of Eq. (\ref{rho_bulk_inter2}) we notice that, thanks to the identity $\cos^2 x = 1/2 + \cos{(2 x)}/2$, one can replace $\left[g_{N p}(z/\sqrt{q})\right]^2$, given in Eq.~(\ref{Plancherel_2}), in the integral over $q$ in (\ref{rho_bulk_inter2}) by $1/2$ (the remaining cosine being highly oscillating for large $N$ and thus subleading). If one finally performs the change of variable $q \to q + z^2$ in (\ref{rho_bulk_inter2}), one obtains finally
\bea\label{rho_bulk_final}
\rho_N\left(x = z/\sqrt{\beta m \omega^2/2} \right) = \frac{\alpha}{\pi \sqrt{y} \sqrt{2 N}}  
\int_0^\infty dq \frac{1}{e^{q+z^2} (e^y -1)^{-1} +1} 
 \frac{1}{\sqrt{q}} = - \frac{\alpha}{\sqrt{2 N \pi y }} {\rm Li}_{1/2}(- (e^y-1) e^{-z^2}) \;,
\eea 
where ${\rm Li}_n(x) = \sum_{k=1}^\infty x^k/k^n$ is the polylogarithm function. Hence from Eq. (\ref{rho_bulk_final}) one obtains the scaling form given in Eq. (7) in the Letter:
\bea
\rho_N(x) \sim \frac{\alpha}{\sqrt{N}} R\left(y = \beta N \hbar \omega, z= x \sqrt{\beta m \omega^2/2} \right)
\eea
with
\bea\label{scaling_function}
R(y,z) = - \frac{1}{\sqrt{2 \pi y}} {\rm Li}_{1/2}(- (e^y-1) e^{-z^2}) \;,
\eea
as given in Eq. (8) of the Letter. 

To analyze the $T \to \infty$ and the $T \to 0$ limits of $\rho_N(x)$ in Eqs. (\ref{rho_bulk_final}) and (\ref{scaling_function}), we need the following asymptotic behaviors of the polylogarithm function:
\begin{eqnarray}\label{asympt_small}
{\rm Li}_{1/2}(X) \sim X \:, \; X \to 0 \;,
\end{eqnarray}
and
\bea\label{asympt_large}
{\rm Li}_{1/2}(-e^X) \sim - \frac{2}{\sqrt{\pi}} X^{1/2} \;, \; X \to \infty \;.
\eea
From these behaviors in (\ref{asympt_small}) and (\ref{asympt_large}), one finds the asymptotic behaviors of the scaling function $R(y,z)$ in (\ref{scaling_function}):
\begin{eqnarray}\label{asympt_scaling}
R(y,z) \sim
\begin{cases}
&\sqrt{\dfrac{y}{2 \pi}} e^{-z^2} \;, \; y \to 0 \\
& \\
& \dfrac{\sqrt{2}}{\pi} \sqrt{1 - \dfrac{z^2}{y}} \;, \; y \to \infty \;, z \to \infty \;{\rm with} \; z^2/y \; {\rm fixed} \;.
\end{cases}
\end{eqnarray}
From the first line of Eq. (\ref{asympt_scaling}), one recovers the $T \to \infty$ limit where the density converges to the Gibbs-Boltzmann (Gaussian) form, given in Eq. (9) of our Letter. On the other hand, from the second line of Eq. (\ref{asympt_scaling}), one obtains the $T \to 0$ limit of the density, which is given by the Wigner semi-circle, given in Eq. (2) of our Letter.

\section{Higer order correlation function and kernel}

Here we focus on the $n$-point correlation function $R_n(x_1,\cdots,x_n)$ defined as
\begin{eqnarray}\label{def_correl_sup}
&&R_n(x_1, \cdots, x_n) = \frac{N!}{(N-n)!} \int_{-\infty}^\infty dx_{n+1} \cdots \int_{-\infty}^\infty dx_{N} P_{\rm joint}(x_1, \cdots, x_n, x_{n+1}, \cdots, x_N) \;,
\end{eqnarray}
where $P_{\rm joint}(x_1, \cdots, x_N)$ is the joint PDF of the $N$ fermions at finite temperature 
in Eq. (\ref{p_start}). To compute the multidimensional integral in Eq. (\ref{def_correl_sup}), it is useful, 
as in the preceding computation of $\rho_N(x)$, to introduce the occupation numbers $m_k = 0,1$. 
Then we note that we can rewrite the norm of any N-body eigenstate appearing in 
(\ref{p_start}), of energy $E$, and corresponding to occupied states $k_1<..<k_N$ as:
\bea
|\Psi_E(x_1,..x_N)|^2 := \frac{1}{N!} \left[\det_{1\leq i,j \leq N} (\varphi_{k_i}(x_j)) \right]^2 = 
\det_{1\leq i,j \leq n} {\tilde K}_N(x_i,x_j; \{m_k\})
\label{kernel.0}
\eea 
where $E=\sum_{k} m_k\, \epsilon_k $ and the kernel $\tilde K_N(x,x';\{m_k \})$ is given by
\begin{eqnarray}\label{def_kernel_canonical}
\tilde K_N(x,x';\{m_k \}) = \sum_{j=1}^N \varphi_{k_j}(x) \varphi_{k_j}(x') =
\sum_{k=0}^\infty m_k \varphi_k(x) \varphi_k(x') \;,
\end{eqnarray}
where $\varphi_k(x)$ is given in Eq. (\ref{def_phik}) and the $m_k=\sum_{j=1}^N \delta_{k,k_j}$  
are $1$ for occupied states and zero otherwise. Now one can integrate $|\Psi_E(x_1,..x_N)|^2$ 
in Eq. (\ref{kernel.0}), for
each fixed $E$, 
over the coordinates $x_{n+1},..x_N$. The 
form of the kernel in Eq. (\ref{def_kernel_canonical}) and the orthonormality of the single particle
wave functions of the $N$ occupied states gaurantees that the determinantal structure 
in Eq. (\ref{kernel.0}) survives the $(N-n)$-fold integration. 
These manipulations are identical to those performed in the context of random matrix 
theory~\cite{mehta_supp,forrester_supp,fyodorov_supp}. There 
they are most often applied to the kernel (\ref{def_kernel_canonical}) built with the $N$ lowest 
energy eigenstates (corresponding
to the ground-state wave function at $T=0$), but the same properties hold for any excited state 
built with $N$ single particle states. 
We thus obtain:
\begin{eqnarray}\label{Rn_occupation}
R_n(x_1, \cdots, x_n) = \frac{1}{Z_N(\beta)} \sum_{\{m_k \}} \left[ \det_{1\leq i,j \leq n} {\tilde K}_N(x_i,x_j; \{m_k\}) e^{-\beta \sum_{k\geq 0} m_k \epsilon_k} \delta\left(\sum_{k\geq 0} m_k,N\right) \right] \;. \label{R1}
\end{eqnarray}

The averaging over the occupation 
numbers in Eq. (\ref{R1}) can be carried out by effectively passing to the grand canonical
ensemble by writing (and thus rendering the $m_k$'s effectively independent Bernoulli random variables)
\begin{equation}
\delta\left(\sum_{k\geq 0} m_k,N\right) = \int_{0}^{2 \pi} {d\lambda\over 2\pi}
\exp\left[ i\lambda\left(\sum_{k\geq 0} m_k-N\right) \right],
\end{equation}
to evaluate both the numerator and the denominator of Eq. (\ref{R1}). 
Now we use the following property:
\bea
\left\langle \det_{1\leq i,j \leq n} \left[ \sum_{k=0}^\infty m_k \varphi_k(x_i) \varphi_k(x_j)\right]\right
\rangle =
\det_{1\leq i,j \leq n} \left[\sum_{k=0}^\infty \langle m_k \rangle \varphi_k(x_i) \varphi_k(x_j) \right]
\eea 
which holds for arbitrary averaging $\langle \ldots \rangle$ for which the variables $m_k$ are independent. It 
is easily seen using the (generalized) Andreief formula \cite{Andreief_supp}:
\bea
\det_{1\leq i,j \leq n} \left[ \sum_{k=0}^\infty m_k \varphi_k(x_i) \varphi_k(x_j)\right] 
= \sum_{k_1<..k_N} m_{k_1} ..m_{k_N} \left(\det_{1\leq i,j \leq n}\left[\varphi_{k_i}(x_j)\right]\right)^2
\eea 
and averaging it, using $\langle m_{k_1} ..m_{k_N} \rangle=\langle m_{k_1} \rangle 
\ldots \langle m_{k_N}\rangle$ (see also \cite{hough_supp} in the context of determinantal processes). 
Here we will use it for an average at 
fixed $\lambda$, i.e., $\langle \ldots \rangle \equiv \langle \ldots \rangle_\lambda$ and 
perform the integral over $\lambda$ at the end. Using the effective independence of the 
$m_k$ in this representation, we thus obtain:
\begin{equation} 
R_n(x_1, \cdots, x_n) = {{ \int_{0}^{2 \pi}  {d\lambda\over 2\pi}
 \det_{1\leq i,j \leq n} {\tilde K}_N(x_i,x_j; {1\over 1+e^{\beta \epsilon_k -i\lambda} })\, 
e^{-\beta J({i\lambda\over \beta})-i\lambda N} }\over  
\int_{0}^{2 \pi}  {d\lambda\over 2\pi} e^{-\beta J({i\lambda\over \beta})-i\lambda N} }\label{newR}
 \end{equation}
 where 
 \begin{equation}
 J(\mu) = -{1\over \beta}\sum_k \ln(1+ e^{-\beta \epsilon_k + \beta\mu})
 \end{equation}
is the grand potential of the grand canonical ensemble. Up to this point, Eq. (\ref{newR}) 
is exact for arbitrary $N$ and $n$. 
The analysis presented in the Letter is then based on the evaluation of the $\lambda$ 
integrals in the denominator and numerator by the saddle point method in the limit of large $N$, 
with the saddle point occurring, in our notation, at $\lambda = \lambda_{sp} =-i\beta \mu$ in both integrals, 
where
$\mu$ is related to $N$ as $N= -\partial J/\partial \mu$,  which is simply  Eq. (\ref{chemical}). This is correct
as long as the corresponding quantity that is averaged does not have the form 
$\exp(cN^a)$ where $a\ge1$ \cite{comment}. Said otherwise, the quantity that is 
computed should not change the value of the chemical potential in the saddle point calculation. 

Therefore, in the large $N$ limit, one obtains from~(\ref{newR}):
\begin{eqnarray}\label{det_process}
R_n(x_1, \cdots, x_n) \simeq \det_{1 \leq i,j \leq n} K_N(x_i,x_j) \;,
\end{eqnarray} 
where the kernel $K_N(x,x')$, obtained from (\ref{def_kernel_canonical}) by replacing $m_k$ by its grand canonical average, i.e. its average at the saddle point $\langle m_k \rangle := \langle m_k \rangle_{\lambda_{sp}}$ in (\ref{fermi_factor}), is given by
\bea\label{kernel_final}
K_N(x,x') = \sum_{k=0}^\infty \frac{\varphi_k(x) \varphi_k(x')}{e^{\beta(\epsilon_k - \mu)} + 1} \;,
\eea
while the chemical potential $\mu$ is fixed by Eq. (\ref{chemical}).  
The case $n=1$ then yields the result for the density in Eq. (\ref{dens1}). 
We remark that the quantity $R_n$ is referred to as the correlation density in the mathematics literature; 
this is because the quantity $R_n(x_1, \cdots, x_n) dx_1\cdots dx_n$ is the probability that there is a 
particle in each of the intervals $[x_i,x_i+dx_i]$, $1\leq i\leq n$. Clearly such quantities exist and make 
sense for ensembles where the particle number varies. Indeed, note that if one formulates the problem 
from the start in the grand canonical ensemble, then Eq. (\ref{det_process}) is exact as an equality. Hence 
the kernel $K_N(x_i,x_j)$ {\it exactly} describes the statistics of a system in the grand canonical 
ensemble at the chemical potential $\mu$ corresponding to $N$ for all values of $\mu$, and not only those 
corresponding to $N$ large (see also Ref. \cite{Joh07_supp}).

These results in Eqs. (\ref{det_process}) and (\ref{kernel_final}) are given in the Letter in Eqs. (25) and 
(26). We now analyze the kernel $K_N(x,x')$ in the bulk limit. The analysis of $K_N(x,x')$ at the edge is 
outlined in the Letter.

\subsection*{Asymptotic limit in the bulk}

We analyze the kernel $K_N(x,x')$ in the bulk where both $x = u N^{-1/2}/\alpha$ and $x' = u' N^{-1/2}/\alpha$ are close to the center of the trap (and $x-x'$ is of the order of the typical inter-particle distance). Hence we analyze the formula (\ref{kernel_final}) and (\ref{chemical}) $N \to \infty$, $\beta \to 0$ keeping $y = \beta N \hbar \omega$ in (\ref{def_yz}) fixed. In this limit the chemical potential $\mu$ is given by Eq. (\ref{mu_bulk}) and the large $N$ analysis of $K_N(x,x')$ (\ref{kernel_final}) can be performed along the same lines as we did before for the density $\rho_N(x)$ yielding eventually Eq. (\ref{rho_bulk_inter2}). Indeed, using the Plancherel-Rotach asymptotic expansions (\ref{Plancherel_1}) and (\ref{Plancherel_2}) we obtain:
\bea\label{kernel_bulk_inter}
K_N(x = u N^{-1/2}/\alpha, x' = u' N^{-1/2}/\alpha) \simeq \frac{\alpha \sqrt{2 N}}{\pi} \int_{0}^\infty \frac{dp}{\sqrt{p}} \frac{1}{e^{y\,p} (e^y -1)^{-1} +1} 
  g_{N p}\left(\frac{u}{N \sqrt{p}}\right) g_{N p}\left(\frac{u'}{N \sqrt{p}}\right) \;.
\eea
From the explicit expression of $g_N(X)$ in Eq. (\ref{Plancherel_2}), one obtains straighforwardly
\bea\label{simplif_gNp}
g_{N p}\left(\frac{u}{N \sqrt{p}}\right) \simeq \cos{\left(2 u \sqrt{p} - N p \frac{\pi}{2} \right)} \;.
\eea
And therefore the product $g_{N p}\left(\frac{u}{N \sqrt{p}}\right) g_{N p}\left(\frac{u'}{N \sqrt{p}}\right)$ in Eq. (\ref{kernel_bulk_inter})
 reads, for large $N$:
\bea\label{prod_g}
g_{N p}\left(\frac{u}{N \sqrt{p}}\right) g_{N p}\left(\frac{u'}{N \sqrt{p}}\right) \simeq \frac{1}{2} \cos{\left(2 \sqrt{p}(u-u') \right)} + \frac{1}{2} \cos{\left(2 \sqrt{p}(u+u') - N p \pi \right)}  \;.
\eea 
The second term in Eq. (\ref{prod_g}) is highly oscillating in the large $N$ limit and hence the leading contribution, once inserted in the integral over $p$ in Eq. (\ref{kernel_bulk_inter}), comes from the first term of Eq. (\ref{prod_g}), which is independent of $N$. Therefore we finally obtain
\bea
K_N(x = u N^{-1/2}/\alpha, x' = u' N^{-1/2}/\alpha) \sim \alpha \sqrt{N} \frac{1}{\pi\sqrt{2}} \int_0^\infty \frac{dp}{\sqrt{p}} \frac{\cos(2 \sqrt{p}(u-u'))}{e^{yp}(e^y-1)^{-1}+1} \;.
\eea
Finally, performing the change of variable $p \to p/y$, one obtains the formula given in Eq. (27) in the Letter (see also Refs. \cite{Verba_supp,Joh07_supp} for alternative derivations of this kernel).


\begin{thebibliography}{100}


\bibitem{BDZ08}
I. Bloch, J. Dalibard, and W. Zwerger, Rev. Mod. Phys. {\bf 80}, 885 (2008).

\bibitem{GPS08}
S. Giorgini, L. P. Pitaevski, and S. Stringari, Rev. Mod. Phys. {\bf 80}, 1215 (2008).



\bibitem{calabrese_prl} P. Calabrese, M. Mintchev, and E. Vicari, Phys. Rev. Lett. {\bf 107}, 020601 (2011); J. Stat. Mech. P09028 (2011). 

\bibitem{vicari_pra} E. Vicari, Phys. Rev. A {\bf 85}, 062104 (2012).

\bibitem{vicari_pra2} M. Campostrini and E. Vicari, Phys. Rev. A {\bf 82}, 063636 (2010).

\bibitem{vicari_pra3} A. Angelone, M. Campostrini, and E. Vicari, Phys. Rev. A {\bf 89}, 023635 (2014).

\bibitem{eisler_prl}
V. Eisler, Phys. Rev. Lett. {\bf 111}, 080402 (2013).

\bibitem{marino_prl}
R. Marino, S. N. Majumdar, G. Schehr, P. Vivo, Phys. Rev. Lett. {\bf 112}, 254101 (2014).

\bibitem{CDM14}
P. Calabrese, P. Le Doussal, S. N. Majumdar, preprint arXiv:1411.4421.

\bibitem{mehta}{M.~L. Mehta, {\it Random Matrices} (Academic Press, Boston, 1991).}

\bibitem{forrester}{ P.~J. Forrester, {\it Log-Gases and Random Matrices} (London Mathematical Society monographs, 2010).  }


\bibitem{TW_GUE}
C.~A. Tracy and H.~Widom, Commun. Math. Phys. \textbf{159}, 151 (1994).

\bibitem{fredholm}
We recall that, for a trace-class operator $K(x,y)$ such that ${\rm Tr} K = \int dx K(x,x)$ is well defined, 
$\det(I - K) = \exp{[-\sum_{n=1}^\infty{{\rm Tr \,} K^n}/{n}]}$, where ${\rm Tr}\, K^n = \int dx_1 \cdots \int dx_n K(x_1,x_2) K(x_2,x_3)\cdots K(x_n,x_1)$. The effect of the projector $P_\xi$ in (\ref{eq:F2}) is simply to restrict the integrals over $x_i$ to the interval $[\xi,+\infty )$.



\bibitem{For93}
P. J. Forrester, Nucl. Phys. B {\bf 402}(3), 709 (1993).

\bibitem{MNS94} M. Moshe, H. Neuberger, B. Shapiro, Phys. Rev. Lett. {\bf 73}, 1497 (1994).



\bibitem{Verba}
A. M. Garc{\' i}a-Garc{\' i}a, J.~J.~M.~ Verbaarschot, Phys. Rev. E {\bf 67}, 046104 (2003). 


\bibitem{JP1}
{R. Allez, J.~-P. Bouchaud, A. Guionnet, Phys. Rev. Lett. {\bf 109}, 094102 (2012).}

\bibitem{JP2}
{R. Allez, J.~-P. Bouchaud, S. N. Majumdar, P. Vivo, J. Phys. A: Math. Theor. {\bf 46}, 015001 (2013).}



\bibitem{Joh07}
{K. Johansson, Probab. Theory Rel. {\bf 138}, 75 (2007).}

\bibitem{ISS13}
T. Imamura, T. Sasamoto, H. Spohn, J. Phys. A: Math. Theor. {\bf 46} 355002 (2013).



\bibitem{footnote} 
{$\gamma=2^{2/3} \lambda$ in terms of $\lambda$ defined in 
\cite{CLR10,DOT10}.}

\bibitem{SS10} T. Sasamoto, H. Spohn, Phys. Rev. Lett. {\bf 104}, 230602 
(2010).

\bibitem{CLR10} P. Calabrese, P. Le Doussal, A. Rosso, Europhys. Lett. {\bf 
90}, 20002 (2010).

\bibitem{DOT10} V. Dotsenko, Europhys. Lett. {\bf 90}, 20003 (2010).

\bibitem{ACQ11} G. Amir, I. Corwin, J. Quastel, Comm. Pure and Appl. Math.
{\bf 64}, 466 (2011).





\bibitem{unp} {D. S. Dean, P. Le Doussal, S. N. Majumdar, G. Schehr, see supplemental material, which includes Refs \cite{fyodorov,Andreief,hough}.}


\bibitem{fyodorov}{ Y. V. Fyodorov, {\it Introduction to the Random Matrix Theory: Gaussian Unitary 
Ensemble and Beyond}, Lond. Math. Soc. Lect. Note Ser. {\bf 322}, 31 (2005).} 

\bibitem{Andreief}
C. Andreief, M\'em. de la Soc. Sci., Bordeaux, (3) {\bf 2}, 1 (1883).  


\bibitem{hough} J.~B. Hough, M.~Krishnapur, Y.~Peres and B.~Vir\'ag, Probability Surveys {\bf 3}, 206 (2006).


\bibitem{compm} The analysis at the edge $x =- \sqrt{2N}$ is clearly identical.

\bibitem{FFG06}
P. J. Forrester, N. E. Frankel, T. M. Garoni, J. Math. Phys. {\bf 47}, 023301 (2006).

\bibitem{BB91}
M. Bowick, E. Br\'ezin, Phys. Lett. B {\bf 268}, 21 (1991). 




\bibitem{comR}Clearly $R_1(x) = N \rho_{N}(x,T)$.



\bibitem{johansson} 
See e.g. K. Johansson, {\it Random matrices and determinantal processes}, in Lecture Notes of the Les
Houches Summer School 2005 (A. Bovier, F. Dunlop, A. van Enter,
F. den Hollander, and J. Dalibard, eds.), Elsevier Science, (2006); arXiv:math-ph/0510038.


\bibitem{unpub}
D. S. Dean, P. Le Doussal, S. N. Majumdar, G. Schehr, to be published. 


\end{thebibliography}

\begin{thebibliography}{10}

\bibitem{FFG06_supp}
P. J. Forrester, N. E. Frankel, T. M. Garoni, J. Math. Phys. {\bf 47}, 023301 (2006).

\bibitem{mehta_supp}{M.~L. Mehta, {\it Random Matrices} (Academic Press, Boston, 1991).}

\bibitem{forrester_supp}{ P.~J. Forrester, {\it Log-Gases and Random Matrices} (London Mathematical Society monographs, 2010).  }

\bibitem{fyodorov_supp}{ Y. V. Fyodorov, {\it Introduction to the Random Matrix Theory: Gaussian Unitary 
Ensemble and Beyond} arXiv:math-ph/0412017, Lond.Math.Soc.Lect.Note Ser.322:31-78,2005.} 

\bibitem{Andreief_supp}
C. Andreief, M\'em. de la Soc. Sci., Bordeaux, (3) 2, 1 (1883).  


\bibitem{hough_supp} J.~B. Hough, M. Krishnapur, Y. Peres and B. Vir\'ag, Probability Surveys, {\bf 3}, 206 (2006).

\bibitem{comment} This is clearly the case for the scaling quantities studied here.

\bibitem{Verba_supp}
A. M. Garc{\' i}a-Garc{\' i}a, J.~J.~M.~ Verbaarschot, Phys. Rev. E {\bf 67}, 046104 (2003). 


\bibitem{Joh07_supp}
{K. Johansson, Probab. Theory Rel. {\bf 138}, 75 (2007).}


\end{thebibliography}
\end{document}